\documentclass[]{spie}  

\usepackage{algorithm,algorithmic}
\usepackage{comment}

   \usepackage[pdftex]{graphicx}
  \graphicspath{{Figures/}}
\usepackage{multicol, blindtext}
\usepackage{amsmath}
\usepackage{enumerate}
\usepackage{amsfonts}

\usepackage[]{graphicx}

\title{Kronecker STAP and SAR GMTI} 

\author{Kristjan H. Greenewald\supit{a}, Edmund G. Zelnio\supit{b}, and Alfred O. Hero III\supit{a}
\skiplinehalf
\supit{a}University of Michigan, Ann Arbor MI, USA; \\
\supit{b}Air Force Research Laboratory, Dayton OH, USA
}

\authorinfo{Further author information: (Send correspondence to K. Greenewald)\\K. Greenewald: E-mail: greenewk@umich.edu\\  E. Zelnio: E-mail: edmund.zelnio@us.af.mil \\ A. Hero: E-mail: hero@umich.edu}

\begin{document} 
  \maketitle

 \maketitle
\begin{abstract}

As a high resolution radar imaging modality, SAR detects and localizes non-moving targets accurately, giving it an advantage over lower resolution GMTI radars. Moving target detection is more challenging due to target smearing and masking by clutter.  Space-time adaptive processing (STAP) is often used on multiantenna SAR to remove the stationary clutter and enhance the moving targets. In (Greenewald et al., 2016) \cite{greenewaldAES}, it was shown that the performance of STAP can be improved by modeling the clutter covariance as a space vs. time Kronecker product with low rank factors, providing robustness and reducing the number of training samples required. In this work, we present a massively parallel algorithm for implementing Kronecker product STAP, enabling application to very large SAR datasets (such as the 2006 Gotcha data collection) using GPUs. Finally, we develop an extension of Kronecker STAP that uses information from multiple passes to improve moving target detection. 




\end{abstract}


\section{Introduction}


The detection (and tracking) of moving objects is an important task for scene understanding, as motion often indicates human related activity \cite{newstadt2013moving}. Radar sensors are uniquely suited for this task, as object motion can be discriminated via the Doppler effect. In this work, we propose a spatio-temporal decomposition method of detecting ground based moving objects in airborne Synthetic Aperture Radar (SAR) imagery, also known as SAR GMTI (SAR Ground Moving Target Indication).

Radar moving target detection modalities include MTI radars \cite{newstadt2013moving,ender1999space}, which use a low carrier frequency and high pulse repetition frequency to directly detect Doppler shifts. This approach has significant disadvantages, however, including low spatial resolution, small imaging field of view, and the inability to detect stationary or slowly moving targets. The latter deficiency means that objects that move, stop, and then move are often lost by a tracker.

SAR, on the other hand, typically has extremely high spatial resolution and can be used to image very large areas, e.g. multiple square miles in the Gotcha data collection \cite{GotchaData}. As a result, stationary and slowly moving objects are easily detected and located \cite{ender1999space,newstadt2013moving}. Doppler, however, causes smearing and azimuth displacement of moving objects \cite{jao2001theory}, making them difficult to detect when surrounded by stationary clutter. Increasing the number of pulses (integration time) simply increases the amount of smearing instead of improving detectability \cite{jao2001theory}. Several methods have thus been developed for detecting and potentially refocusing \cite{cristallini2013efficient,cerutti2012optimum} moving targets in clutter. Our goal in \cite{greenewaldAES} and this work is to remove the disadvantages of MTI and SAR by combining their strengths (the ability to detect Doppler shifts and high spatial resolution) using space time adaptive processing (STAP) with a novel Kronecker product spatio-temporal covariance model.  


SAR systems can either be single channel (standard single antenna system) or multichannel. Standard approaches for the single channel scenario include autofocusing \cite{fienup2001detecting} and velocity filters. Autofocusing works only in low clutter, however, since it may focus the clutter instead of the moving target \cite{fienup2001detecting,newstadt2013moving}. Velocity filterbank approaches used in track-before-detect processing \cite{jao2001theory} involve searching over a large velocity/acceleration space, which often makes computational complexity excessively high. Attempts to reduce the computational complexity have been proposed, e.g. via compressive sensing based dictionary approaches \cite{khwaja2011applications} and Bayesian inference \cite{newstadt2013moving}, but remain computationally intensive.

Multichannel SAR has the potential for greatly improved moving target detection performance \cite{ender1999space,newstadt2013moving}. Standard multiple channel configurations include spatially separated arrays of antennas, flying multiple passes (change detection), using multiple polarizations, or combinations thereof \cite{newstadt2013moving}. 


\subsection{Previous Multichannel Approaches}
Several techniques exist for using multiple radar channels (antennas) to separate the moving targets from the stationary background. SAR GMTI systems have an antenna configuration such that each antenna transmits and receives from approximately the same location but at slightly different times \cite{GotchaData,ender1999space,newstadt2013moving,cerutti2012optimum}. Along track interferometry (ATI) and displaced phase center array (DPCA) are two classical approaches \cite{newstadt2013moving} for detecting moving targets in SAR GMTI data, both of which are applicable only to the two channel scenario. Both ATI and DPCA first form two SAR images, each image formed using the signal from one of the antennas. To detect the moving targets, ATI thresholds the phase difference between the images and DPCA thresholds the magnitude of the difference. A Bayesian approach using a parametric cross channel covariance generalizing ATI/DPCA to $p$ channels was developed in \cite{newstadt2013moving}, and a unstructured method fusing STAP and a test statistic in \cite{cerutti2012optimum}. Space-time Adaptive Processing (STAP) learns a spatio-temporal covariance from clutter training data, and uses these correlations to filter out the stationary clutter while preserving the moving target returns \cite{ender1999space,ginolhac2014exploiting,klemm2002principles}. 


In \cite{greenewaldAES}, we proposed a covariance-based STAP algorithm with a customized Kronecker product covariance structure. The SAR GMTI receiver consists of an array of $p$ phase centers (antennas) processing $q$ pulses in a coherent processing interval. Define the array $\mathbf{X}^{(m)} \in \mathbb{C}^{p\times q}$ such that $X_{ij}^{(m)}$ is the radar return from the $j$th pulse of the $i$th channel in the $m$th range bin. Let $\mathbf{x}_m = \mathrm{vec}(\mathbf{X}^{(m)})$. The target-free radar data $\mathbf{x}_m$ is complex valued and is assumed to have zero mean. Define
\begin{align}
\mathbf{\Sigma} = \mathrm{Cov}[\mathbf{x}] = E[\mathbf{x} \mathbf{x}^H].
\end{align}

The training samples, denoted as the set $\mathcal{S}$, used to estimate the SAR covariance $\mathbf{\Sigma}$ are collected from $n$ representative range bins. 
The standard sample covariance matrix (SCM) is given by
\begin{align}
\label{Eq:SCM}
\mathbf{S} = \frac{1}{n}\sum_{m\in\mathcal{S}} \mathbf{x}_m \mathbf{x}_m^H.
\end{align}
If $n$ is small, $\mathbf{S}$ may be rank deficient or ill-conditioned \cite{newstadt2013moving,ginolhac2014exploiting,greenewaldArxiv,greenewaldSSP2014}, and it can be shown that using the SCM directly for STAP requires a number $n$ of training samples that is at least twice the dimension $pq$ of $\mathbf{S}$ \cite{reed1974rapid}. In this data rich case, STAP performs well \cite{newstadt2013moving,ender1999space,ginolhac2014exploiting}. However, with $p$ antennas and $q$ time samples (pulses), the dimension $pq$ of the covariance is often very large, making it difficult to obtain a sufficient number of target-free training samples. This so-called ``small $n$ large $pq$" problem leads to severe instability and overfitting errors, compromising STAP tracking performance.

By introducing structure and/or sparsity into the covariance matrix, the number of parameters and the number of samples required to estimate them can be reduced. 
As the spatiotemporal clutter covariance $\mathbf{\Sigma}$ is low rank \cite{brennan1992subclutter,ginolhac2014exploiting,rangaswamy2004robust,ender1999space}, Low Rank STAP (LR-STAP) clutter cancelation estimates a low rank clutter subspace from $\mathbf{S}$ and uses it to estimate and remove the rank $r$ clutter component in the data \cite{bazi2005unsupervised,ginolhac2014exploiting}, reducing the number of parameters from $O(p^2q^2)$ to $O(rpq)$. 
Efficient algorithms, including some involving subspace tracking, have been proposed \cite{belkacemi2006fast,shen2009reduced}. Other methods adding structural constraints such as persymmetry \cite{ginolhac2014exploiting,conte2003exploiting}, and robustification to outliers either via exploitation of the SIRV model \cite{ginolhac2009spatio} or adaptive weighting of the training data \cite{gerlach2011robust} have been proposed. Fast approaches based on techniques such as Krylov subspace methods \cite{goldstein1998multistage,honig2002adaptive,pados2007short,scharf2008subspace} and adaptive filtering \cite{rui2011reduced,rui2010reduced} exist. All of these techniques remain sensitive to outlier or moving target corruption of the training data, and generally still require large training sample sizes \cite{newstadt2013moving}. 


Instead, for SAR GMTI \cite{greenewaldAES} proposed exploiting the explicit space-time arrangement of the covariance by modeling the clutter covariance matrix $\mathbf{\Sigma}_c$ as the Kronecker product of two smaller matrices 
\begin{equation}
\label{KronApprox}
\mathbf{\Sigma}_c = \mathbf{A}\otimes \mathbf{B},
\end{equation}
where $\mathbf{A} \in \mathbb{C}^{p\times p}$ is rank 1 and $\mathbf{B}\in \mathbb{C}^{q\times q}$ is low rank.
In this setting, the $\mathbf{B}$ matrix is the ``temporal (pulse) covariance" and $\mathbf{A}$ is the ``spatial (antenna) covariance," both determined up to a multiplicative constant. We note that this model is not appropriate for classical GMTI STAP, as in that phased array configuration the clutter covariance has a different spatio-temporal structure that is not separable, arising from the angle-dependent changes in the phase pattern. 



In practical SAR applications, it is of interest to be able to efficiently and accurately perform STAP on very large scenes. In this work, we propose an iterative parallel algorithm to estimate the low-rank Kronecker factors of the clutter covariance. We are then able to implement the method on a GPU, gaining very significant speed-ups relative to the CPU only implementation \cite{greenewaldAES}. We envision this allowing Kron STAP to be viable on large-scale systems. 

Additionally, as in the Gotcha data collection \cite{GotchaData}, it is common for data to be collected on multiple passes, i.e. the radar flies past a scene on the same path at different times. This allows change detection to be performed by (in effect) differencing the two images. When such data is available, much of the stationary clutter will not change between passes. We propose to exploit this fact in Kron STAP, and propose a new method for doing so. Significant clutter cancelation gains are observed, enabling better enhancement of moving targets.


To summarize, the main contributions of this paper are: 1) the design and implementation of an efficient, massively parallel algorithm for estimating the clutter subspace in the Kron STAP framework; 2) a method for extending Kron STAP to multipass (change detection) problems; and 3) real data results on the full 2006 Gotcha dataset, demonstrating the scalability and further demonstrating the advantages of our method. 

The remainder of the paper is organized as follows. Section \ref{Sec:Model}, presents the multichannel SIRV radar model. An extension to the case of moving target detection with multiple passes is presented in Section \ref{Sec:MultiPass}. Our massively parallel low rank Kronecker product covariance estimation algorithm is given in Section \ref{Sec:KSTAP}, and we review the Kronecker STAP filters in Section \ref{Sec:STAP}.
Section \ref{Sec:Results} gives simulation results and applies our algorithms to the Gotcha dataset and Section \ref{Sec:Conc} concludes the paper. 

In this work, we denote vectors as lower case bold letters, matrices as upper case bold letters, the complex conjugate as $a^*$, the matrix Hermitian as $\mathbf{A}^H$, and the Hadamard (elementwise) product as $\mathbf{A}\odot \mathbf{B}$.


\section{SIRV Clutter Model}
\label{Sec:Model}
In this section, we review the multichannel SAR clutter model \cite{greenewaldAES}. Let $\mathbf{X}\in \mathbb{C}^{p \times q}$ be an array of radar returns from an observed range bin across $p$ channels and $q$ pulses. We model $\mathbf{x} = \mathrm{vec}(\mathbf{X}^T)$ as a spherically invariant random vector (SIRV) with the following decomposition \cite{yao1973representation,rangaswamy2004robust,ginolhac2014exploiting,ginholhac2013performance}: 
\begin{align}
\label{Eq:decomp}
\mathbf{x} = \mathbf{x}_{target} + \mathbf{x}_{clutter} + \mathbf{x}_{noise} = \mathbf{x}_{target} + \mathbf{n},
\end{align}
where $\mathbf{x}_{noise}$ is Gaussian sensor noise with $\mathrm{Cov}[\mathbf{x}_{noise}]=\sigma^2 \mathbf{I} \in \mathbb{C}^{pq \times pq}$ and we define $\mathbf{n} =\mathbf{x}_{clutter} + \mathbf{x}_{noise}$. The signal of interest $\mathbf{x}_{target}$ is the sum of the spatio-temporal returns from all moving objects, modeled as non-random, in the range bin. The return from the stationary clutter is given by $\mathbf{x}_{clutter}  =\tau \mathbf{c}$ where $\tau$ is a random positive scalar having arbitrary distribution, known as the \emph{texture}, and $\mathbf{c} \in \mathbb{C}^{pq}$ is a multivariate complex Gaussian distributed random vector, known as the \emph{speckle}. We define $\mathrm{Cov}[\mathbf{c}]= \mathbf{\Sigma}_c$. 
The means of the clutter and noise components of $\mathbf{x}$ are zero. The resulting clutter plus noise ($\mathbf{x}_{target} = 0$) covariance is given by
\begin{align}
\label{Eq:Cov}
\mathbf{\Sigma} = E[\mathbf{n}\mathbf{n}^H] =  E[\tau^2]\mathbf{\Sigma}_c + \sigma^2 \mathbf{I}.
\end{align}
The ideal (no calibration errors) random speckle $\mathbf{c}$ is of the form \cite{newstadt2013moving,ender1999space,cerutti2012optimum}
\begin{align}
\label{Eq:7}
\mathbf{c} = \mathbf{1}_p \otimes \tilde{\mathbf{c}},
\end{align}
where $\tilde{\mathbf{c}} \in \mathbb{C}^q$. The representation \eqref{Eq:7} follows because the antenna configuration in SAR GMTI is such that the $k$th antenna receives signals emitted at different times at approximately (but not necessarily exactly) the same point in space \cite{newstadt2013moving,GotchaData}. This is achieved by arranging the $p$ antennas in a line parallel to the flight path, and delaying the $k$th antenna's transmission until it reaches the point $x_i$ in space associated with the $i$th pulse. The representation \eqref{Eq:7} gives a clutter covariance of 
\begin{align}
\label{Eq:ClutterCov}
\mathbf{\Sigma}_c = \mathbf{11}^T \otimes \mathbf{B}, \quad \quad \mathbf{B} = E[\tilde{\mathbf{c}} \tilde{\mathbf{ c }}^H],
\end{align}
where $\mathbf{B}$ 
depends on the spatial characteristics of the clutter in the region of interest and the SAR collection geometry \cite{ender1999space}. 
While in SAR GMTI $\mathbf{B}$ is not exactly low rank, it is approximately low rank in the sense that significant energy concentration in a few principal components is observed over small regions \cite{greenewaldAES,borcea2013synthetic}.


Due to the long integration time and high cross range resolution associated with SAR, the returns from the general class of moving targets are more complicated, making simple Doppler filtering difficult. During short intervals for which targets have constant Doppler shift $f$ (proportional to the target radial velocity) within a range bin, the return has the form
\begin{align}
\label{Eq:Moving}
\mathbf{x} = \alpha \mathbf{d} = \alpha \mathbf{a}(f) \otimes \mathbf{b}(f),
\end{align}
where $\alpha$ is the target's amplitude, $\mathbf{a}(f) = [\begin{array}{cccc} 1 & e^{j2\pi \theta_1(f)} & \dots & e^{j 2 \pi \theta_{p}(f)} \end{array}]^T$, the $\theta_i$ depend on Doppler shift $f$ and the platform speed and antenna separation \cite{newstadt2013moving}, and $\mathbf{b} \in \mathbb{C}^q$ depends on the target, $f$, and its cross range path. The unit norm vector $\mathbf{d} =  \mathbf{a}(f) \otimes \mathbf{b}(f)$ is known as the \emph{steering vector}. For sufficiently large $\theta_i(f)$, $\mathbf{a}(f)^H \mathbf{1}$ will be small and the target will lie outside of the SAR clutter spatial subspace. The overall target return can be approximated as a series of constant-Doppler returns, hence the overall return should lie outside of the clutter spatial subspace.  Furthermore, as observed in \cite{fienup2001detecting,greenewaldAES}, for long integration times the return of a moving target is significantly different from that of uniform stationary clutter, implying that moving targets generally lie outside the temporal clutter subspace \cite{fienup2001detecting} $\mathbf{U}_B \in \mathbb{C}^{q \times r_b}$ as well. 

In practice, the signals from each antenna have gain and phase calibration errors that vary slowly across angle and range \cite{newstadt2013moving}. It was shown in \cite{newstadt2013moving} that in SAR GMTI these calibration errors can be accurately modeled as constant over small regions. Let the calibration error on antenna $i$ be $h_i e^{j\phi_i}$ and $\mathbf{h} = [\begin{array}{ccc} h_1 e^{j\phi_1},& \dots, & h_p e^{j\phi_p}\end{array}]$, giving an observed return $\mathbf{x}' = (\mathbf{h} \otimes \mathbf{I})\odot \mathbf{x}$ and a clutter covariance of 
\begin{align}
\label{Eq:KronCov}
\tilde{\mathbf{\Sigma}}_c  = (\mathbf{h} \mathbf{h}^H) \otimes \mathbf{B}= \mathbf{A} \otimes \mathbf{B}
\end{align}
implying that the $\mathbf{A}$ in \eqref{KronApprox} has rank $r_a = 1$, implying the existence of a spatial clutter subspace spanned by $\mathbf{U}_A \in \mathbb{C}^{p \times 1}$.

Since moving targets lie outside the clutter subspaces (spanned by $\mathbf{U}_A$ and $\mathbf{U}_B$ in space and time respectively), estimating the clutter subspace (covariance) and projecting it away via STAP (Section \ref{Sec:STAP}) will improve target detection performance.

\section{Multipass Clutter Model}
\label{Sec:MultiPass}

In surveillance applications, it is often of interest to determine what, if anything, has changed in a scene between a reference time $t_0$ and a later time $t_1$, e.g. disappearance/appearance of parked vehicles, or the appearance of vehicle footprints \cite{newstadt2013moving,bazi2005unsupervised,bovolo2005detail,ranney2006signal}. When SAR is used for such change detection applications, the radar platform will generally fly past the scene and form a ``reference'' image at time $t_0$, and then at time $t_1 > t_0$ fly a path as close as possible to the original and form a new ``mission'' image. These images are then compared and changes detected. However, moving targets will almost always be detected as changes, along with the changes in the stationary scene background \cite{newstadt2013moving}. When changes of background are of primary interest, moving targets may in fact mask changes in the stationary scene due to displacement and smearing. Hence, it is advantageous to identify moving targets in both scenes prior to or parallel to background change detection. In addition, it may be of interest to detect moving targets in the imagery for their own sake \cite{newstadt2013moving}. We thus exploit the additional scene information arising from having two images to better estimate the clutter subspace. 

We consider the general case of $K$ passes. To form the model, we concatenate the spatial channels of all $K$ registered clutter phase histories ($\mathbf{X}_k \in \mathbb{C}^{p \times q}$), forming a ``$Kp$ channel phase history"
\begin{align}
\mathbf{X} = \left[\begin{array}{c}\mathbf{X}_1\\\vdots \\ \mathbf{X}_K\end{array}\right] \in \mathbb{C}^{Kp \times q}.
\end{align}
Now from the previous section, it is clear that $\mathrm{Cov}[\mathrm{vec}(\mathbf{X}_k^T)] = \mathbf{A}\otimes \mathbf{B}$, implying the columns of $\mathbf{X}_k$ are spanned by $\mathbf{U}_A \in \mathbb{C}^{p \times 1}$ and its rows are spanned by $\mathbf{U}_B \in \mathbb{C}^{q \times r_B}$. 

As a result, every column of $\mathbf{X}$ can be written as 
\begin{equation}
\left[\begin{array}{c} c_1\mathbf{U}_{A_1} \\ \vdots \\  c_K\mathbf{U}_{A_K} \end{array} \right]
\end{equation}
for some scalars $c_k$. Thus the columns of $\mathbf{X}$ lie in the $K$-dimensional subspace spanned by $\tilde{\mathbf{U}}_A = [\mathbf{U}_{A_1}, \dots, \mathbf{U}_{A_K}]$. The rows of $\mathbf X$ are all spanned by $\mathbf{U}_B$ so the overall clutter vector $\mathbf{x} = \mathrm{vec}(\mathbf{X}^T)$ exists in the subspace spanned by
\begin{equation}
\tilde{\mathbf{U}} = \tilde{\mathbf{U}}_A \otimes \mathbf{U}_B,
\end{equation}
giving a clutter covariance of the form
\begin{equation}
\mathrm{Cov}[\mathbf{x}] = \tilde{\mathbf \Sigma}_c = \tilde{\mathbf{A}} \otimes \tilde{\mathbf{B}}.
\end{equation}
The factors $\tilde{\mathbf{A}} \in \mathbb{C}^{Kp \times Kp}$ and $\tilde{\mathbf{B}} \in \mathbb{C}^{q \times q}$ are rank $K$ and rank $r_b$ respectively.
Thus, a rank $K$ spatial clutter subspace and a low rank temporal subspace can be estimated using LR-Kron.


\section{Parallel Kronecker Subspace Estimation}
\label{Sec:KSTAP}
\label{Sec:Alg}

In \cite{greenewaldAES} we developed a subspace estimation algorithm that accounts for spatio-temporal covariance structure and has low computational complexity. In this work, we present a modified algorithm that best exploits the massively parallel structure available on GPUs.

As in \cite{greenewaldAES} we fit a low rank Kronecker product model to the sample covariance matrix ${\mathbf{ S}}$. Specifically, we minimized the Frobenius norm of the residual errors in the approximation of $\mathbf S$ by the low rank Kronecker model \eqref{Eq:KronCov}, subject to $\mathrm{rank}(\mathbf{A}) \leq r_a, \mathrm{rank}(\mathbf{B}) \leq r_b$, where the goal is to estimate $E[\tau^2] \mathbf{\Sigma}_c$. The optimal estimates of the Kronecker matrix factors $\mathbf{A}$ and $ \mathbf{B}$ in \eqref{Eq:KronCov} are given by
\begin{equation}
\label{Eq:SparseOpt}
\hat{\mathbf{A}},\hat{\mathbf{B}} = \arg\min_{\mathrm{rank}({\mathbf{A}}) \leq r_a,\mathrm{rank}({\mathbf{B}}) \leq r_b}\| \mathbf{S}-{ \mathbf{A}}\otimes{\mathbf{B}}\|_F^2.
\end{equation}

The minimization \eqref{Eq:SparseOpt} will be simplified by using the patterned block structure of $\mathbf A \otimes \mathbf B$. In particular, for a $pq \times pq$ matrix $\mathbf{M}$, define $\{\mathbf{M}(i,j)\}_{i,j=1}^p$ to be its $q \times q$ block submatrices, i.e. $\mathbf{M}(i,j) = [\mathbf{M}]_{(i-1)q + 1:iq,(j-1)q+1:jq}$. Also, let $\overline{\mathbf{M}} = \mathbf{K}_{p,q}^T \mathbf{M} \mathbf{K}_{p,q}$ where $\mathbf{K}_{p,q}$ is the $pq \times pq$ permutation operator such that $\mathbf{K}_{p,q} \mathrm{vec}(\mathbf{N}) = \mathrm{vec}(\mathbf{N}^T)$ for any $p\times q$ matrix $\mathbf{N}$.

The invertible Pitsianis-VanLoan rearrangement operator $\mathcal{R}(\cdot)$ maps $pq\times pq$ matrices to $p^2 \times q^2$ matrices and, as defined in \cite{tsiliArxiv,werner2008estimation} sets the $(i-1)p + j$th row of $\mathcal{R}(\mathbf{M})$ equal to $\mathrm{vec}(\mathbf{M}(i,j))^T$, i.e. 
\begin{align}
\label{Eq:SVD}
\mathcal{R}(\mathbf{M}) &= [\begin{array}{ccc} \mathbf{m}_1 & \dots & \mathbf{m}_{p^2}\end{array}]^T,\\\nonumber
\mathbf{m}_{(i-1)p+j} &= \mathrm{vec}(\mathbf{M}(i,j)), \quad i,j = 1,\dots,p.
\end{align}
In Algorithm \ref{alg:Re}, we show our pseudocode implementation of the rearrangement operator on the GPU, where rounding down and up are denoted by $\mathrm{floor}(\cdot)$ and $\mathrm{ceil}(\cdot)$ respectively.

Applying the rearrangement operator to \eqref{Eq:SparseOpt}, we obtain
\begin{equation}
\label{Eq:SparseOptGPU}
\hat{\mathbf{a}},\hat{\mathbf{b}} = \arg\min_{\mathrm{rank}({\mathbf{A}}) \leq r_a,\mathrm{rank}({\mathbf{B}}) \leq r_b}\| \mathcal{R}(\mathbf{S})-{ \mathbf{a}}{\mathbf{b}^T}\|_F^2
\end{equation}
where $\mathbf{a} = \mathrm{vec}(\mathbf{A}), \mathbf{b} = \mathrm{vec}(\mathbf{B})$.





With the low rank constraints, there is no closed-form solution of \eqref{Eq:SparseOpt}. An iterative alternating minimization was derived in \cite{greenewaldAES}, but involved a eigendecomposition of a $q \times q$ matrix at each step. Eigendecomposition requires $O(q^3)$ operations, and in practical applications $q \gg p$, making this problematic to do at every iteration.

Our alternative method is summarized by Algorithm \ref{alg:LRKron}. In Algorithm \ref{alg:LRKron}, $\mathrm{EIG}_{r}(\mathbf{M})$ denotes the matrix obtained by truncating the Hermitian matrix $\mathbf M$ to its first $r$ principal components, i.e.
\begin{equation}
\mathrm{EIG}_r(\mathbf{M}) := \sum_{i=1}^r \sigma_i \mathbf{u}_i \mathbf{u}_i^H,
\end{equation}
where $\sum_i \sigma_i \mathbf{u}_i \mathbf{u}_i^H$ is the eigendecomposition of $\mathbf{M}$, and the (real and positive) eigenvalues $\sigma_i$ are indexed in order of decreasing magnitude. To avoid the $q \times q$ eigendecomposition, we approximate the solution of \eqref{Eq:SparseOptGPU} by solving the \eqref{Eq:SparseOptGPU} with $(r_a = r_a, r_b = q)$ and then projecting the resulting estimate of $\mathbf{B}$ down to rank $r_b$. As a result, eigendecomposition is performed only once (as in classical LR-STAP). The eigendecompositions of the estimate of $\mathbf{A}$ are of size $p \times p$ and thus are computationally trivial in most applications. Note that every other step in the iteration is a simple matrix multiplication, making acceleration on the GPU highly practical overall.

We initialize the algorithm by setting 
\begin{equation}
\mathbf{a}_0 = \frac{1}{q^2} \mathbf{R}\mathbf{1}_{q \times 1}
\end{equation}
implying
\begin{equation}
\mathbf{A}_0 = \mathrm{vec}^{-1}(\mathbf{a}_0) = \frac{1}{q^2}(\mathbf{I}_p \otimes \mathbf{1}_{q \times 1})^T\mathbf{\Sigma}_{SCM} (\mathbf{I}_p \otimes \mathbf{1}_{q \times 1})
\end{equation}
which is positive semidefinite Hermitian.
 


Algorithm \ref{alg:LRKron} is called low rank Kronecker product covariance estimation, or LR-Kron. It is shown  in \cite{greenewaldAES} that since the initialization $\mathbf{A}_0$ is positive semidefinite Hermitian the LR-Kron estimator $\hat{\mathbf{A}}\otimes \hat{\mathbf{B}}$ is positive semidefinite Hermitian and is thus a valid covariance matrix of rank $r_a r_b$.

\begin{algorithm}[H]
\caption{Rearrangement Operator}
\label{alg:Re}
\begin{algorithmic}[1]
\STATE Input $p,q,$ $\mathbf{S}\in \mathbb{C}^{pq \times pq}$
\STATE $\mathbf{s} \gets \mathrm{vec}(\mathbf{S})$
\STATE $\mathbf{P} \gets [1, \dots, q^2]^T \mathbf{1}_{1 \times p^2}$
\STATE $\mathbf{J} \gets \mathbf{1}_{q^2 \times 1}[1, \dots, p^2] $
\FOR{$i = 1:q^2, j = 1:p^2$}
\STATE $L_{ij} \gets P_{ij} - q \cdot \mathrm{floor}((P_{ij}-1)/q) + q(J_{ij} - p \cdot \mathrm{floor}((J_{ij}-1)/p - 1) $
\STATE $N_{ij} \gets \mathrm{ceil}(P_{ij}/q) + q \cdot (\mathrm{ceil}(J_{ij}/p)-1)$
\STATE $M_{ij} \gets L_{ij} + (N_{ij}-1)pq$
\STATE $R_{ji} \gets s_{M_{ij}}$
\ENDFOR
\RETURN $\mathbf R \in \mathbb{C}^{p^2 \times q^2}$.
\end{algorithmic}
\end{algorithm}

\begin{algorithm}[H]
\caption{LR-Kron Covariance Estimation}
\label{alg:LRKron}
\begin{algorithmic}[1]
\STATE $\mathbf{S} = \mathbf{\Sigma}_{SCM} = \frac{1}{n}\sum_n \mathbf{x}_n\mathbf{x}_n^T$, tolerance $\epsilon$, $p \ll q$.
\STATE $\mathbf{R} \gets \mathcal{R}(\mathbf{\Sigma}_{SCM})$ using Alg. \ref{alg:Re}.
\STATE Initialize $\mathbf{a}_0 = \frac{1}{q^2}\mathbf{R}\mathbf{1}_{q \times 1}$, $k = 0$.
\WHILE{$\eta_{k} - \eta_{k-1} > \epsilon$} 
\STATE $\mathbf{b}_k \gets \frac{1}{\|\mathbf{a}_k\|_2^2} \mathbf{R}^T \mathbf{a}_k^*$.
\STATE $\mathbf{a}_{k+1} \gets \frac{1}{\|\mathbf{b}_k\|_2^2} \mathbf{R}\mathbf{b}_{k}^*$.
\STATE $\mathbf{a}_{k+1} \gets \mathrm{vec}(\mathrm{EIG}_{r_a}(\mathrm{vec}^{-1}(\mathbf{a}_{k+1})))$.
\STATE $\eta_{k+1} \gets \frac{\| \mathbf{R}-\mathbf{a}_{k+1} \mathbf{b}_{k}^T\|_F}{\| \mathbf{R}\|_F}$.

\STATE $k \gets k + 1$.
\ENDWHILE
\STATE $\hat{\mathbf{A}} \gets \mathrm{vec}^{-1}({\mathbf{a}_k})$.
\STATE $\hat{\mathbf{B}} \gets \mathrm{EIG}_{r_b} (\mathrm{vec}^{-1}({\mathbf{b}_k}))$.
\RETURN $\hat{\mathbf{A}} ,\hat{\mathbf{B}}$.

\end{algorithmic}
\end{algorithm}

\section{Kronecker STAP Filters}
\label{Sec:STAP}


In this section, we present our method for applying our low rank Kronecker clutter covariance estimate to STAP. Let the vector $\mathbf{d}$ be a spatio-temporal ``steering vector" \cite{ginolhac2014exploiting}, that is, a matched filter for a specific target location/motion profile. For a measured array output vector $\mathbf x$  define the STAP filter output $y={\mathbf w}^T {\mathbf x}$, where  $\mathbf w$ is a vector of spatio-temporal filter coefficients. By \eqref{Eq:decomp} and \eqref{Eq:Moving} we have
\begin{align}
\label{Eq:Breakdown}
y = \mathbf{w}^H\mathbf{x} = \alpha \mathbf{w}^H \mathbf{d}+ \mathbf{w}^H \mathbf{n}.
\end{align}

The goal of STAP is to design the filter $\mathbf{w}$ such that the clutter is canceled ($\mathbf{w}^H \mathbf{n}$ is small) and the target signal is preserved ($\mathbf{w}^H \mathbf{d}$ is large).

For a given target with spatio-temporal steering vector $\mathbf{d}$, this is quantified as the SINR (signal to interference plus noise ratio), defined as the ratio of the power of the filtered signal $\alpha \mathbf{w}^H \mathbf{d}$ to the power of the filtered clutter and noise \cite{ginolhac2014exploiting}
\begin{align}
\label{Eq:SINRdef}
\mathrm{SINR}_{out} = \frac{|\alpha|^2 |\mathbf{w}^H\mathbf{d}|^2}{E[\mathbf{w}^H \mathbf{n}\mathbf{n}^H \mathbf{w}]} = \frac{|\alpha|^2 |\mathbf{w}^H\mathbf{d}|^2}{\mathbf{w}^H \mathbf{\Sigma} \mathbf{w}},
\end{align}
where $\mathbf{\Sigma}$ is the clutter plus noise covariance in \eqref{Eq:Cov}. 

It can be shown \cite{ender1999space,ginolhac2014exploiting} that, if the clutter covariance is known, under the SIRV model the optimal filter at steering vector $\mathbf{d}$ is given by
\begin{align}
\label{Eq:opt}
\mathbf{w} = \mathbf{F}_{opt}\mathbf{d},\quad \mathbf{F}_{opt} = \mathbf{\Sigma}^{-1}.
\end{align}
Since the true covariance is unknown, \cite{greenewaldAES} considered filters of the form
\begin{align}
\label{Eq:GenFilt}
\mathbf{w} = \mathbf{F}\mathbf{d}.
\end{align}

The classical low-rank STAP filter is the projection matrix: 
\begin{align}
\label{Eq:KronUS}
\mathbf{F}_{classical} = \mathbf{I} - \mathbf{U}_A \mathbf{U}_A^H\otimes {\mathbf{U}_B \mathbf{U}_B^H},
\end{align}
where $\mathbf{U}_A, \mathbf{U}_B$ are orthogonal bases for the rank $r_a$ and $r_b$ subspaces of the low rank estimates $\hat{ \mathbf{A}}$ and $\hat {\mathbf{B}}$, respectively, obtained by applying Algorithm \ref{alg:LRKron}. This is the Kronecker product equivalent of the standard STAP projector \eqref{Eq:RegStap}, though it should be noted that \eqref{Eq:KronUS} will require less training data for equivalent performance due to the assumed structure. 

The classical low-rank filter $\mathbf{F} = \mathbf{I} - \mathbf{U}\mathbf{U}^H$ is merely an approximation to the SINR optimal filter $\mathbf{F} = \mathbf{\Sigma}^{-1}$. It was noted in \cite{greenewaldAES}, however, that this may not be the only possible approximation. In particular, the inverse of a Kronecker product is the Kronecker product of the inverses, i.e. $\mathbf{A}\otimes\mathbf{B} = \mathbf{A}^{-1} \otimes \mathbf{B}^{-1}$. Hence, using the low rank filter approximation on $\hat{ \mathbf{A}}^{-1}$ and $\hat{\mathbf{B}}^{-1}$ directly was proposed in \cite{greenewaldAES}. The resulting approximation to $\mathbf{F}_{opt}$ is
\begin{align}
\label{Eq:KronSTAP}
\mathbf{F}_{KSTAP} = (\mathbf{I} - \mathbf{U}_A \mathbf{U}_A^H) \otimes (\mathbf{I} - \mathbf{U}_B \mathbf{U}_B^H) = \mathbf{F}_A \otimes \mathbf{F}_B.
\end{align}
Kron-STAP \cite{greenewaldAES} denotes the method using LR-Kron to estimate the covariance and \eqref{Eq:KronSTAP} to filter the data. This alternative approximation has significant appeal. Note that it projects away both the spatial and temporal clutter subspaces, instead of only the joint spatio-temporal subspace. This is appealing because by \eqref{Eq:Moving}, no moving target should lie in the same spatial subspace as the clutter, and, as noted in Section \ref{Sec:Model}, if the dimension of the clutter temporal subspace is sufficiently small relative to the dimension $q$ of the entire temporal space, moving targets will have temporal factors ($\mathbf{b}$) whose projection onto the clutter temporal subspace are small. Note that in the event $r_b$ is very close to $q$, either truncating $r_b$ to a smaller value (e.g., determined by cross validation) or setting $\mathbf{U}_B = 0$ is recommended to avoid canceling both clutter and moving targets.

Finally, since moving targets do not match the single Kronecker product covariance structure of the clutter, Kron-STAP is shown to be highly robust to uncoordinated moving targets being included in the training data \cite{greenewaldAES}. This is a significant advantage over classical LR-STAP, which is highly vulnerable to such corruption.



\section{Numerical Results}
\label{Sec:Results}

\subsection{Example Timing Simulations}
In this section, we compare wall clock times for LR-Kron covariance estimation on the GPU and on the CPU. The system we used for the comparison has 7 NVIDIA Tesla GPUs, and a 40 core Intel Xeon CPU system. Due to the number of cores, the CPU implementation still exploits parallelism, but still to a much smaller degree than the GPU.

We set $\mathbf{A}$ to be a complex random $p \times p$ rank-one matrix ($r_a = 1$), and $\mathbf{B}$ to be the identity ($r_b = q$). $n=5$ training examples were used, and the timing results were averaged over 10 random trials. For each simulation, we fixed $p$ and varied $q$. 

\begin{figure}[htb]
\centering
\includegraphics[width=3in]{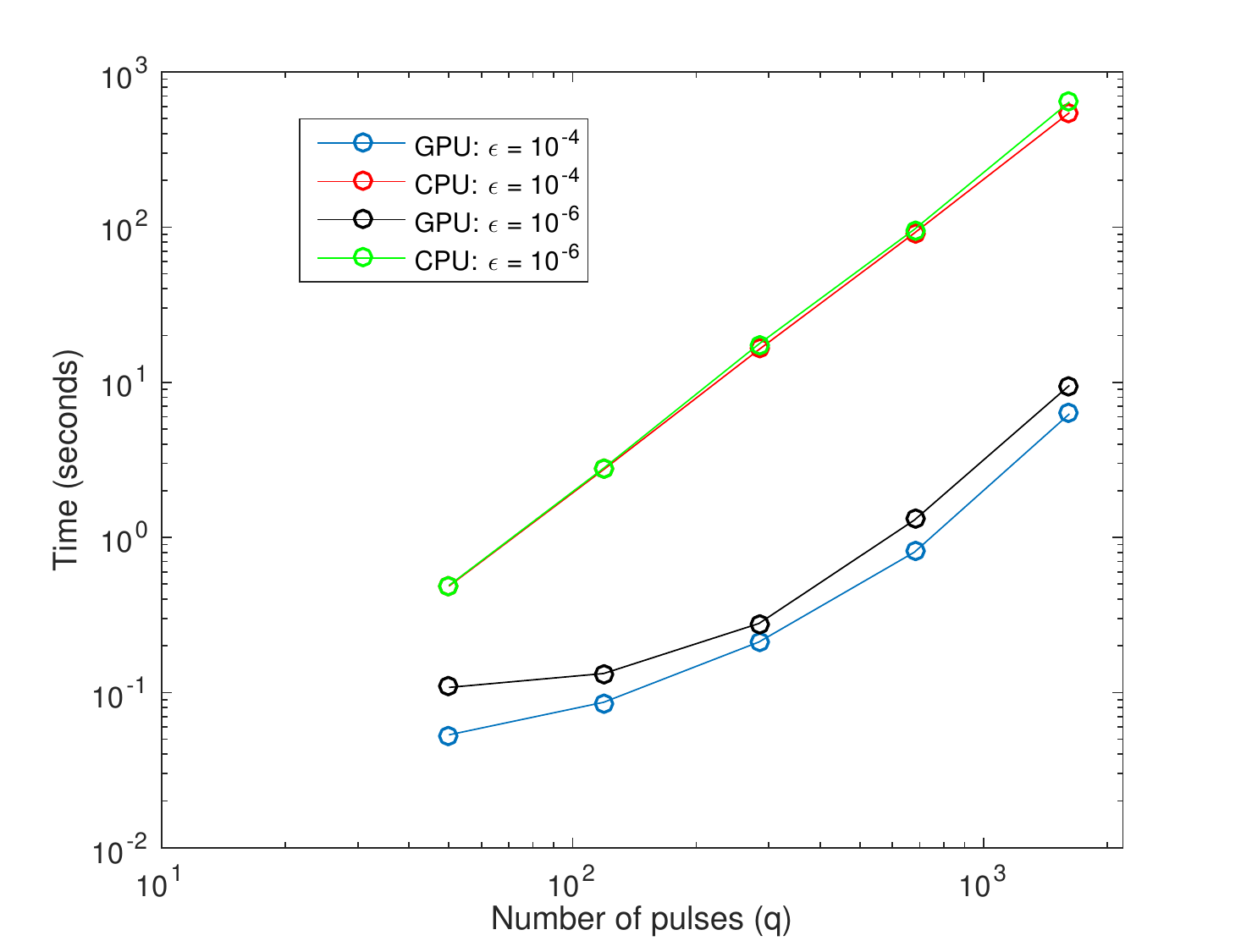}\includegraphics[width=3in]{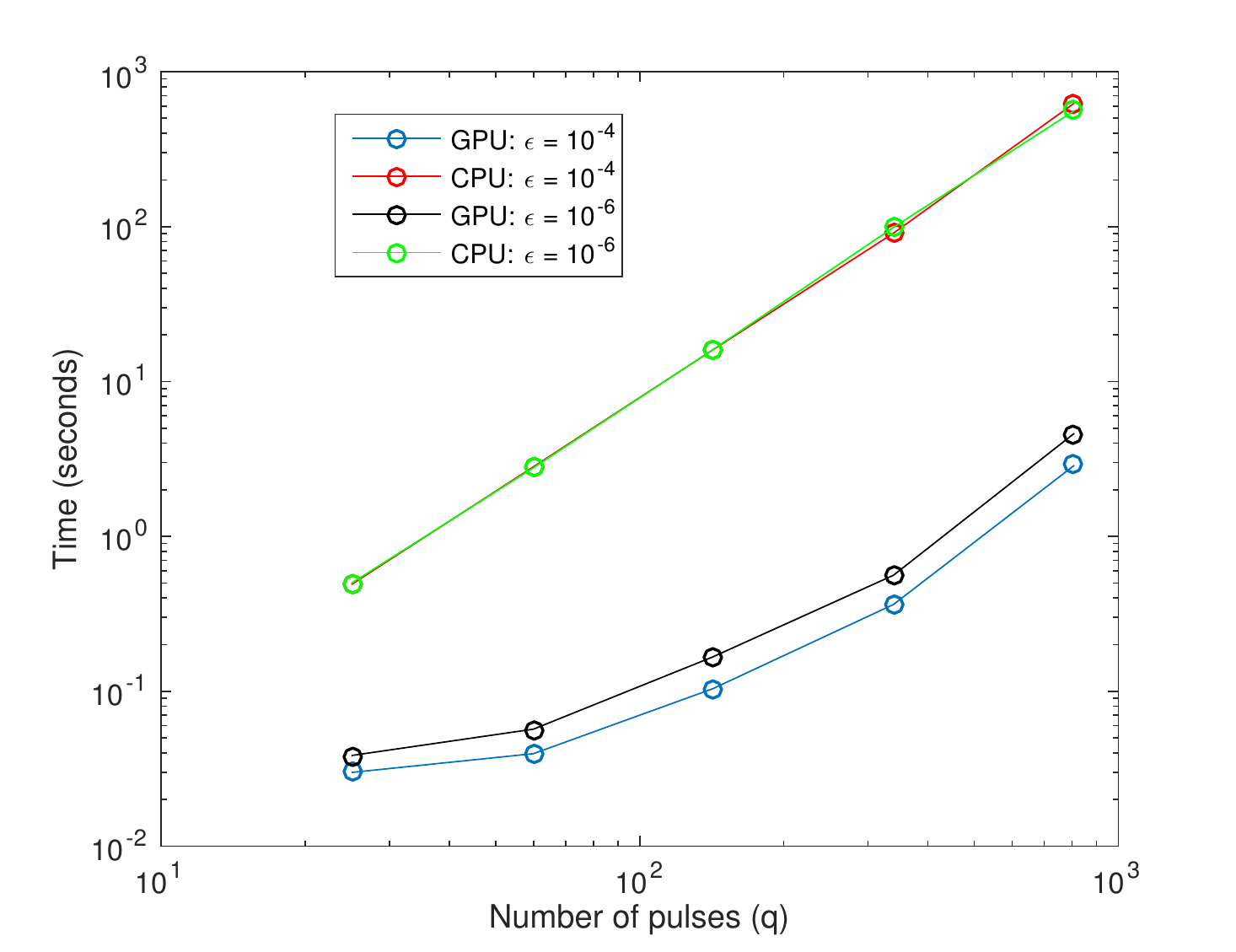}
\caption{Wall clock computation time for LR-Kron covariance estimation on the CPU and GPU for medium ($\epsilon = 10^{-4}$) and low ($\epsilon = 10^{-6}$) tolerances. Left: $p = 3$, right $p = 6$. Note the GPU speedups possible due to massive parallelization. }
\label{fig:time3}
\end{figure}

Figure \ref{fig:time3} (left) shows the results for $p = 3$, $q \in [50, 2500]$ and a medium tolerance ($\epsilon = 10^{-4}$), and for a very low tolerance ($\epsilon = 10^{-6}$). Similar results for $p = 6$, $q \in [25,1250]$ are shown on the right side of Figure \ref{fig:time3}. Note that the CPU time does not increase much (proportionally) with decreasing $\epsilon$, indicating that the rearrangement operation takes a significant portion of the time. The ability of the GPU to handle this operation in parallel gives a significant advantage. 

As expected, the massively parallel structure of the steps of the algorithm allows for orders of magnitude speedup over the same algorithm on the CPU. This implies that very large covariances (large $q$) can be estimated given sufficient memory and GPU resources.


\subsection{Dataset}



For evaluation of the proposed Kron STAP methods, we use measured data from the 2006 Gotcha SAR sensor collection \cite{GotchaData}. This dataset consists of SAR passes through a circular path around a large scene containing multiple roadways and various moving and stationary civilian vehicles. The example images shown in the figures are formed using the backprojection algorithm with Blackman-Harris windowing as in \cite{newstadt2013moving}. For our experiments, we use 31 seconds of data, divided into 1 second (2171 pulse) coherent integration intervals.

As there is no ground truth for all targets in the Gotcha imagery, target detection performance cannot be objectively quantified by ROC curves. 
We thus rely on anecdotal evidence and refer to \cite{greenewaldAES} for a more quantitative analysis of Kron-STAP.

\subsection{Gotcha Experimental Data}



In this subsection, STAP is applied to the Gotcha dataset. For each range bin we construct steering vectors $\mathbf{d}_i$ corresponding to 2000 cross range pixels. In single antenna SAR imagery, each cross range pixel is a Doppler frequency bin that corresponds to the cross range location for a stationary target visible at that SAR Doppler frequency, possibly complemented by a moving target that appears in the same bin. Let $\mathbf{D}$ be the matrix of steering vectors for all Doppler (cross range) bins in each range bin. Then the SAR images at each antenna are given by $\tilde{\mathbf{x}} = \mathbf{I}\otimes\mathbf{D}^H\mathbf{x}$ and the STAP output for a spatial steering vector $\mathbf{h}$ and temporal steering $\mathbf{d}_i$ (separable as noted in \eqref{Eq:Moving}) is the scalar
\begin{align}
y_i(\mathbf{h}) = (\mathbf{h}\otimes \mathbf{d}_i)^H \mathbf{F} \mathbf{x}
\end{align}
Due to their high dimensionality, plots for all values of $\mathbf{h}$ and $i$ cannot be shown. Hence, for interpretability we produce images where for each range bin the $i$th pixel is set as $\max_{\mathbf{h}} |y_i(\mathbf{h})|$. More sophisticated detection techniques could invoke priors on $\mathbf{h}$, but we leave this for future work.

\begin{figure}[]
\centering
\includegraphics[width=6.5in]{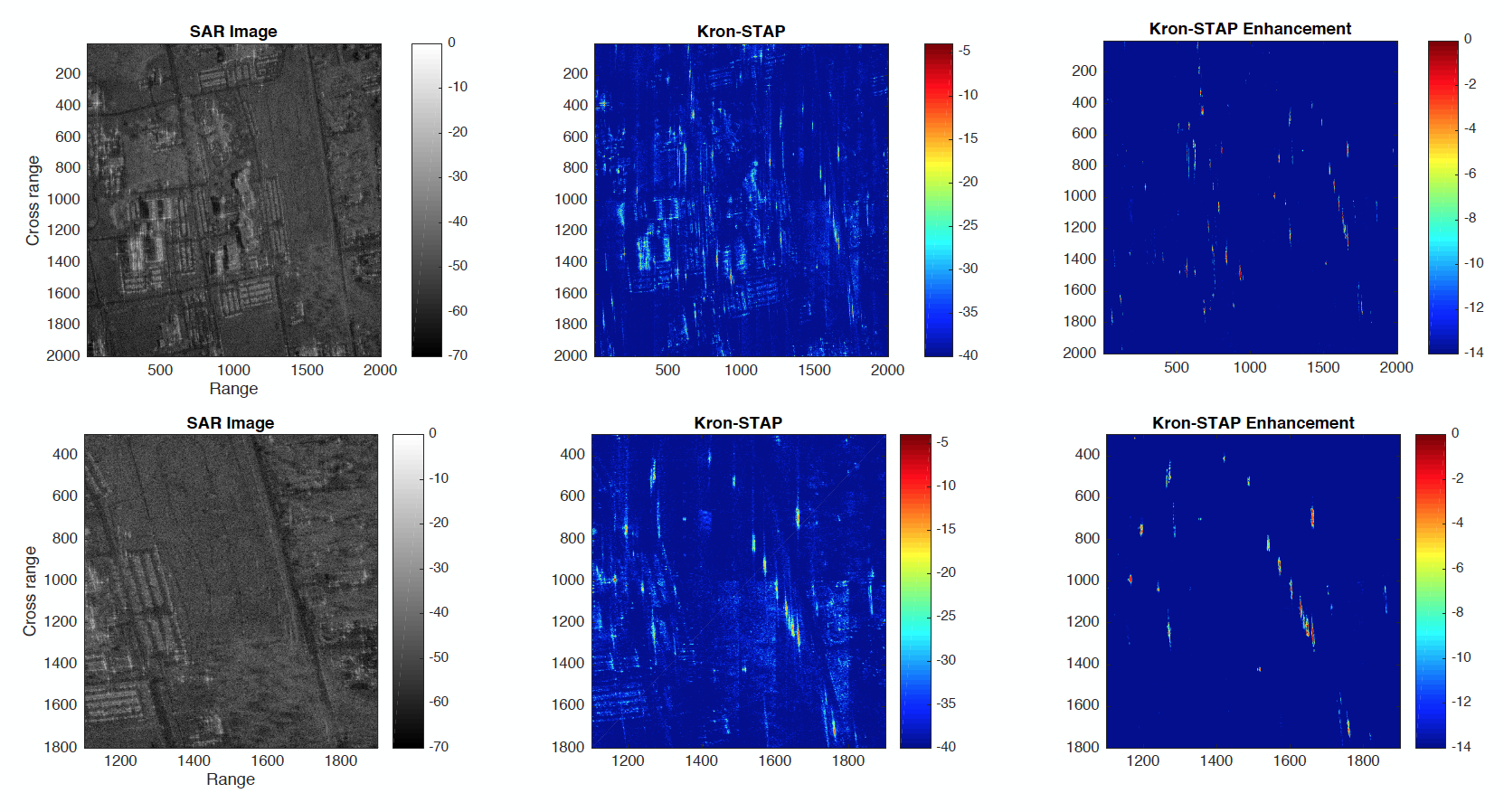}
\caption{Example Kron STAP results on Gotcha scene. Top row: full scene, bottom row: zoomed in on potential example moving targets. For each row the original SAR image is shown with Kron STAP results at each pixel. Note the bright spots in the Kron STAP in the vicinity of roadways, indicating groups of cars. No ground truth is available, however based on movement between adjacent frames most of the bright spots seem to correspond to moving vehicles. } 
\label{Fig:Examples}
\end{figure}

Shown in Figure \ref{Fig:Examples} are results for an examplar SAR scene, showing the original SAR (single antenna) image and the results of spatio-temporal Kronecker STAP. Also shown in the Kron STAP enhancement, which is found by dividing the intensity of the Kron STAP image by the original SAR image. High intensities on the enhancement image indicate that less cancelation has occurred, indicating moving targets. Note the very strong contrast of Kronecker STAP between presumed moving targets and the background. Ground truth is not available, but we used temporally adjacent data to verify that the bright spots in the Kron STAP enhancement image move over time in a manner consistent with moving vehicles. 

\subsection{Multipass Kron STAP}
\begin{figure}[]
\centering
\includegraphics[width=6.5in]{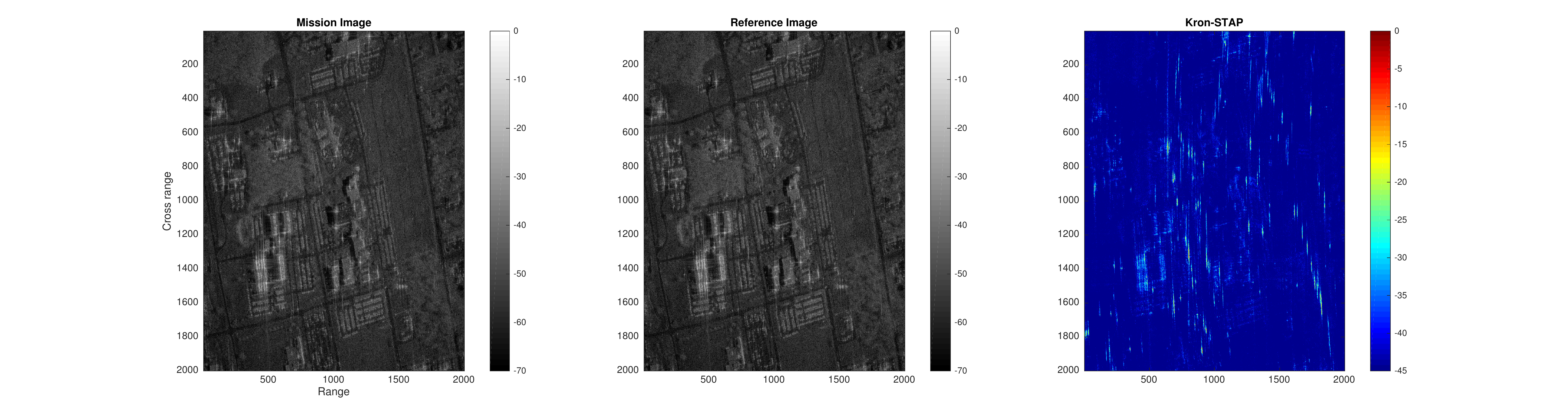}\\
\includegraphics[width=6.5in]{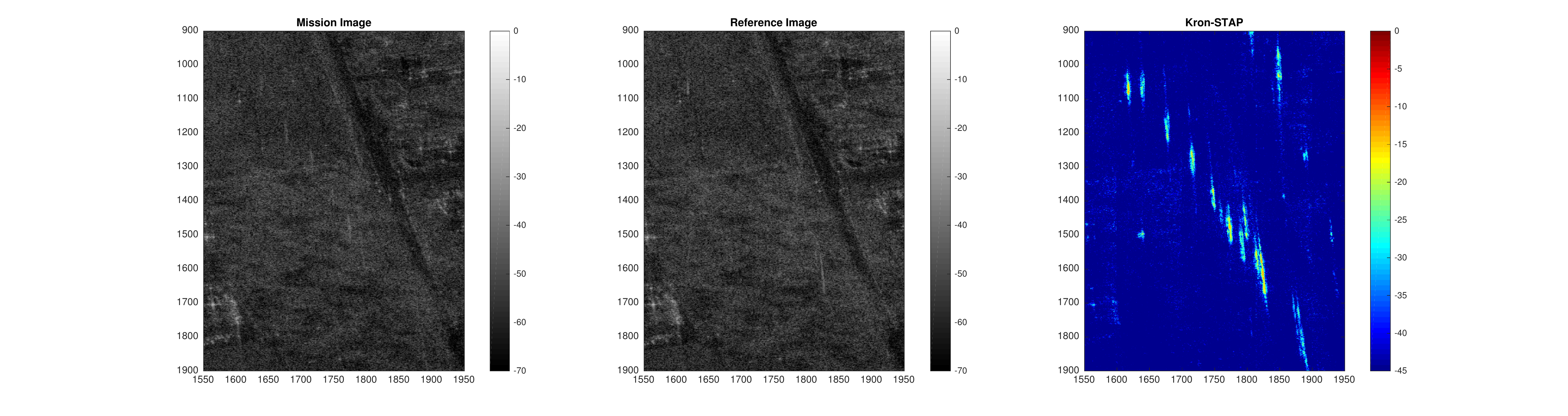}\\
\includegraphics[width=6.5in]{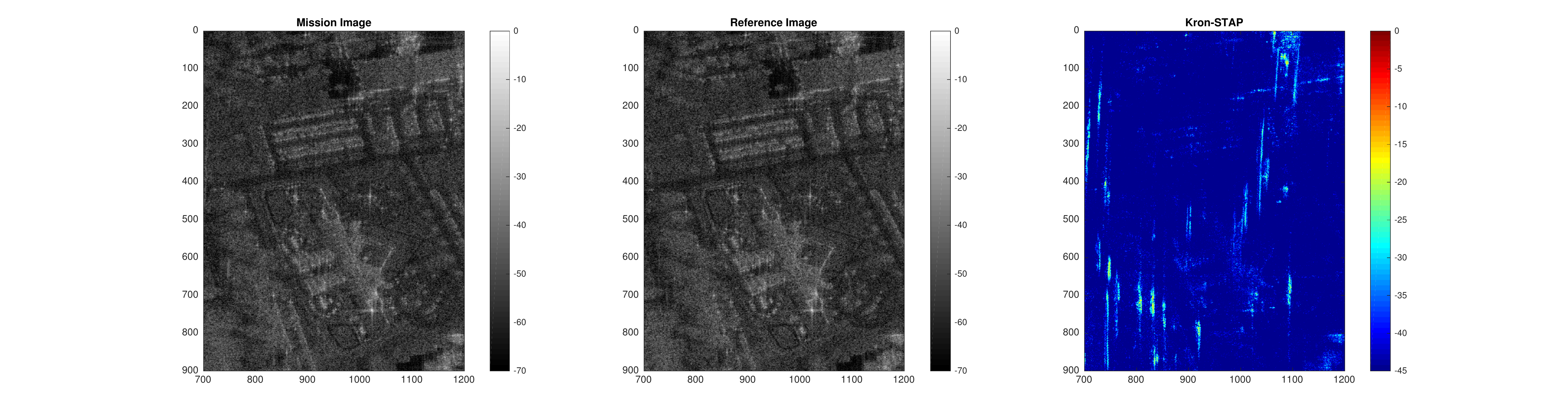}
\caption{Example Multipass Kron STAP. Top row: full scene, bottom two rows: zoomed in on potential example moving targets. For each row the mission and reference images are shown with Multipass Kron STAP. Note the bright spots in the Kron STAP in the vicinity of roadways, indicating groups of cars. No ground truth is available, however based on movement between adjacent frames most of the bright spots seem to correspond to moving vehicles. } 
\label{Fig:Cd}
\end{figure}

Representative two pass Kronecker STAP results followed by noncoherent change detection are shown in Figure \ref{Fig:Cd}. Noncoherent change detection is performed following filtering by reforming each image (via maximum steering vectors as in the previous section) and subtracting the resulting pixel magnitudes. It can be seen that additional clutter cancelation capabilities can be gained by using Kronecker STAP on multiple passes.

\section{Conclusion}
\label{Sec:Conc}






We considered expansions of the Kronecker STAP method in practical directions. Clutter in multiple antenna synthetic aperture radar systems has been shown \cite{greenewaldAES} to have a Kronecker product covariance with low rank factors, and Kronecker STAP was developed to exploit this structure for improved clutter cancelation performance. Kronecker STAP has been proven to improve robustness to corrupted training data, and to dramatically reduce the number of training samples needed. In this work, we developed an implementation of Kronecker STAP implementable on massively parallel architectures such as the GPU, making it scalable to high-dimensional applications. Additionally, we expanded the Kronecker STAP model and method to include multiple passes at different times, enabling improved stationary clutter cancelation in applications where such data is available. Finally, we evaluated our methods on the full 2006 Gotcha dataset, demonstrating computational scalability to such applications and further confirming the clutter cancelation advantages of the Kronecker STAP model and its extensions.

\section{Acknowledgments}
This work was supported in part by Army Research Office grant W911NF-11-1-0391 and Department of the Air Force grant FA8650-15-D-1845. Approved for public release, PA Approval \#88ABW-2016-1820. \

\bibliographystyle{spiebib}
\bibliography{CAMSAP_bib}

\end{document}